\documentclass[reprint,
superscriptaddress,
 amsmath,amssymb,
 aps,pra,
showkeys,
]{revtex4-2}

\usepackage[utf8]{inputenc}
\usepackage{longtable}
\usepackage{morefloats}
\usepackage{tabularx}
\usepackage{multirow}
\usepackage{graphicx}
\usepackage{color}
\usepackage{epsfig,amsbsy}
\usepackage{amsmath,amsfonts,amsthm,amssymb}
\usepackage{appendix}
\usepackage{textcomp}
\usepackage{makeidx}
\usepackage{epstopdf}
\usepackage[normalem]{ulem}
\usepackage[bookmarksnumbered,pdfpagelabels=true,plainpages=false,colorlinks=true,linkcolor=blue,citecolor=red,urlcolor=blue]{hyperref}

\usepackage{xcolor}

\NewDocumentCommand{\fhighlight}{O{blue!40} m m}
{\draw[rounded corners,ultra thin, fill=red,draw=black, opacity=0.1] (m-1-1.north west)rectangle (m-2-2.south east);
\draw[rounded corners,ultra thin, fill=red,draw=black, opacity=0.1] (m-3-3.north west)rectangle (m-4-4.south east);
\draw[rounded corners,ultra thin, fill=blue,draw=black, opacity=0.03] (m-1-3.north west)rectangle (m-2-4.south east);
\draw[rounded corners,ultra thin, fill=blue,draw=black, opacity=0.03] (m-3-1.north west)rectangle (m-4-2.south east);}

\newcommand{\al}[1]{{\color{purple}#1}}
\begin{document}

\title{Synthetic Altermagnetism Beyond the Crystal Limit}

\author{R. A. Gallardo}
\email{rodolfo.gallardo@usm.cl}
\affiliation{Departamento de F\'isica, Universidad T\'ecnica Federico Santa Mar\'ia, Avenida Espa\~na 1680, Valpara\'iso, Chile}

\author{A. M. Le\'on}
\affiliation{Departamento de F\'isica, Facultad de Ciencias, Universidad de Chile, Casilla 653, Santiago, Chile}

\author{J. Lindner}
\affiliation{Helmholtz-Zentrum Dresden-Rossendorf, Institute of Ion Beam Physics and Materials Research, Bautzner Landstr. 400, 01328 Dresden, Germany}

\author{J. W. Gonz\'alez}
\email{jhon.gonzalez@uantof.cl}
\affiliation{Departamento de F\'isica, Universidad de Antofagasta, Av. Angamos 601, Casilla 170, Antofagasta, Chile}

\date{\today }

\begin{abstract}
Altermagnetic magnons in crystalline materials exhibit momentum-dependent splitting whose nodal structure and chiral character are governed by the point-group symmetry of the magnetic sublattice rotation. 
Here, we demonstrate the first synthetic realization of altermagnetic magnonics in a continuum platform composed of antiferromagnetically coupled ferromagnetic films with alternating in-plane exchange anisotropies, showing that the key signatures of altermagnetic magnonics emerge beyond the crystalline setting.
Solving the linearized Landau-Lifshitz equation within a dipole-exchange framework, we show that this architecture reproduces the characteristic momentum-dependent splitting, nodal directions, and anisotropic isofrequency contours of A-type altermagnets.
Long-range dipolar interactions qualitatively reconstruct this exchange-driven spectrum by lifting the nominal nodal degeneracy, hybridizing opposite-chirality modes, and producing a finite, thickness-dependent wave-vector splitting along directions that are nodal in the exchange-only limit. 
Extending the bilayer to finite multilayers reveals that synthetic altermagnetism undergoes a parity-dependent reconstruction that separates surface and bulk altermagnetic excitations.
These results establish altermagnetic magnon phenomenology as an engineerable collective response of dipole-exchange multilayers beyond microscopic crystal symmetries.

\end{abstract}

\maketitle


The recent discovery of altermagnetism has rapidly extended beyond the
electronic sector into the realm of collective spin excitations, giving rise to
the emerging field of altermagnetic magnonics~\cite{Libor23,Liu24,Hoyer25}.
In altermagnetic systems, exchange interactions lift the chirality degeneracy of
magnons, generating momentum-dependent branch splitting and symmetry-enforced
nodal directions absent in conventional
antiferromagnets~\cite{Libor23,Liu24,Hoyer25}.
These distinctive features have attracted considerable attention owing to their potential for field-free spin-wave manipulation, directional magnon transport, chiral magnon filtering, and spin-caloritronic responses~\cite{Markus24,Wu25,Cui26}.
Recent neutron-scattering experiments provide evidence of chiral magnon splitting in candidate altermagnets~\cite{Faure25,Sears26}.
At the same time, these measurements reveal a central challenge: while altermagnetic symmetry permits magnon splitting, its magnitude is governed by microscopic exchange interactions and may become exceedingly small in real materials.
Indeed, neutron scattering has failed to resolve the altermagnetic magnon splitting in MnF$_2$~\cite{morano2025absence}, while in FeF$_2$ the dominant source of magnon splitting is the long-range dipolar interaction rather than altermagnetic exchange~\cite{Sears26}.
These findings establish that the realization of altermagnetic magnonics is not solely a symmetry problem, but also a materials-design challenge requiring platforms with sufficiently large and controllable energy scales for experimental detection and practical implementation.

This challenge motivates the search for synthetic platforms in which altermagnetic symmetry is engineered rather than inherited from a particular crystal structure.
Recent theoretical advances show that the characteristic $d$-wave spin splitting of altermagnets emerges in layered systems through rotated anisotropic hopping parameters~\cite{Asgharpour25}.
In oxide superlattices, interface-driven orbital reconstruction reveals that layer parity and spacer thickness act as key control parameters for stabilizing two-dimensional altermagnetic states~\cite{Xiang26}.
Complementary micromagnetic studies demonstrate that $d$-wave exchange anisotropy can be encoded in mesoscale spin textures, extending altermagnetic symmetry concepts beyond atomistic crystals~\cite{Jiang25}.
Together, these works establish altermagnetism as an engineerable symmetry principle rather than an exclusive property of bulk compounds.
Yet, despite this rapid progress, a synthetic realization of altermagnetic magnonics in a continuum dipole-exchange platform remains absent.

Realizing altermagnetic magnonics in synthetic magnetic heterostructures raises an additional challenge specific to the mesoscopic regime.
Unlike tight-binding or atomistic descriptions, magnetic films and multilayers operate in the continuum dipole-exchange regime, where spin-wave dynamics are governed by the interplay between exchange and long-range magnetostatic interactions~\cite{Camley99,Camley87,Hillebrands90}.
Recent theoretical studies show that dipolar interactions hybridize altermagnetic magnons of opposite chirality and induce anisotropic level repulsion when treated as corrections to the spectrum of crystalline altermagnets~\cite{Jin25}, while surface spin waves provide additional signatures of broken $\mathcal{PT}$ symmetry~\cite{Sun26}.
Whether the characteristic $d$-wave nodal structure of altermagnetic magnons survives or is qualitatively reconstructed when magnetostatic interactions are intrinsic to, rather than perturbative corrections to, the spin-wave dynamics therefore remains an open question.

Here, we demonstrate that the key signatures of altermagnetic magnonics, including momentum-dependent branch splitting, symmetry-enforced nodal directions, and anisotropic isofrequency contours, emerge in a synthetic continuum platform composed of antiferromagnetically coupled ferromagnetic films with alternating in-plane exchange anisotropies.
This architecture generates an effective altermagnetic interaction proportional to $k_x^2-k_z^2$, reproducing the characteristic magnon spectrum of A-type altermagnets without relying on an underlying altermagnetic crystal.
Within a full dipole-exchange framework, we show that long-range magnetostatic interactions do not merely perturb this exchange-driven spectrum but qualitatively reconstruct it by hybridizing opposite-chirality modes and generating finite thickness-dependent wave-vector splitting along directions where exchange symmetry alone predicts exact nodal degeneracies.
Extending the architecture to finite multilayers reveals a parity-dependent reconstruction in which even-layer stacks preserve altermagnetic branch splitting across bulk and surface modes, whereas odd-layer stacks exhibit partial surface suppression associated with uncompensated sublattice termination.
These results establish stack parity and surface termination as geometric control parameters for synthetic altermagnetic magnonics in dipole-exchange multilayers, representing design principles with no direct analogue in bulk
crystalline altermagnets.

\begin{figure*}[t]
\includegraphics[width=0.99\textwidth]{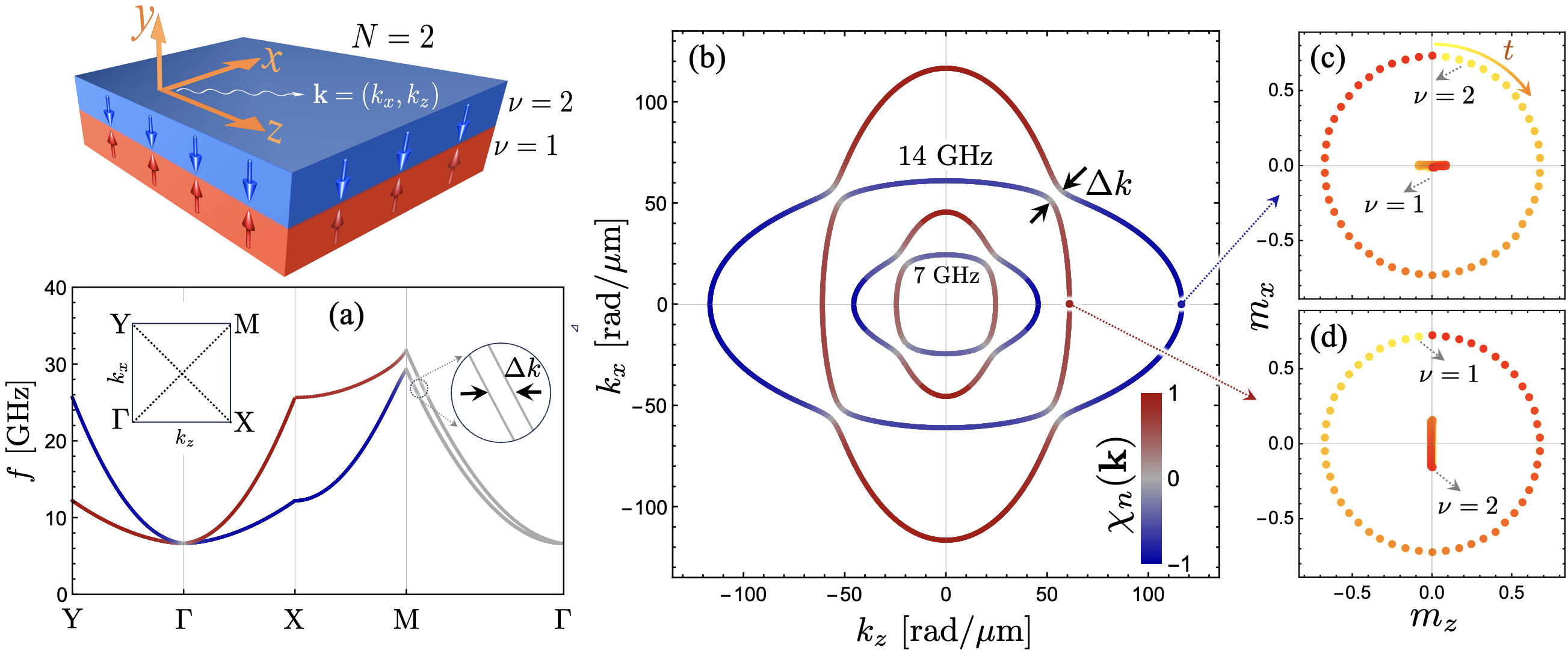}
\caption{Synthetic altermagnetic bilayer composed of two antiferromagnetically coupled ferromagnetic layers ($N=2$) with opposite equilibrium magnetizations. (a) Spin-wave dispersion along the high-symmetry path $\rm Y-\Gamma-X-M-\Gamma$ of the Brillouin zone. The inset illustrates the reciprocal-space directions and the wave-vector splitting $\Delta k$.
(b) Isofrequency contours in the $(k_x,k_z)$ plane for excitation frequencies of 7 and 14 GHz. The contours exhibit a pronounced anisotropic deformation and a momentum-dependent separation of the two magnon branches, characteristic of altermagnetic order. The color scale represents the magnon polarization parameter $\chi$, highlighting the opposite polarization character of the two branches. The arrows indicate the wave-vector splitting $\Delta k$ evaluated along the $\rm \Gamma$--$\rm M$ direction.
(c,d) Time evolution of the normalized dynamic magnetization components $(m_x,m_z)$ for the two magnon branches at $f=14$ GHz and $k_x=0$. The temporal coordinate progresses from light to dark colors (yellow to red), while $\nu=1$ and $\nu=2$ denote the bottom and top magnetic layers, respectively.
}
  \label{FIG1}
\end{figure*}

Our synthetic altermagnetic multilayer consists of $N$ ferromagnetic films stacked along the $y$ direction and laterally extended in the $x$--$z$ plane (see Supplemental Material \cite{SuppMater}). Neighboring layers are antiferromagnetically coupled, leading to the compensated equilibrium configuration $\mathbf{M}^{\rm eq}_{\nu}=(-1)^{\nu+1}M_{\rm s}\hat{\mathbf y}$, where $\nu$ labels the magnetic layer. Spin-wave excitations with in-plane wave vector $\mathbf{k}=(k_x,k_z)$ are described by $\mathbf{M}_{\nu}=\mathbf{M}^{\rm eq}_{\nu}+\mathbf{m}_{\nu}e^{i(k_xx+k_zz-\omega t)}$. Linearization of the Landau--Lifshitz equation then yields the eigenvalue problem
\begin{equation}
i\frac{\omega}{\mu_0\gamma}\mathbf{m}
=
\widetilde{\mathcal{D}}(\mathbf{k})\mathbf{m},
\label{EQdyn}
\end{equation}
where $\mathbf{m}=(m_{x_1},m_{z_1},\ldots,m_{x_N},m_{z_N})^{T}$ contains the dynamic amplitudes of all layers, and $\widetilde{\mathcal{D}}$ includes anisotropic intralayer exchange, interlayer exchange, perpendicular anisotropy, and dipolar interactions. The explicit matrix elements are given in the Supplemental Material \cite{SuppMater}.
The key ingredient is an alternating anisotropic exchange environment. In layer $\nu$, the intralayer exchange field is
\begin{equation}
\mathbf{H}^{\rm ex}_{\nu}
=
\left[\ell^{(\nu)}_{{\rm ex}_x}\right]^2
\frac{\partial^2\mathbf{M}_{\nu}}{\partial x^2}
+
\left[\ell^{(\nu)}_{{\rm ex}_z}\right]^2
\frac{\partial^2\mathbf{M}_{\nu}}{\partial z^2},
\end{equation}
with the principal exchange axes interchanged between adjacent layers. For a bilayer, the resulting exchange anisotropy enters the spectrum through the splitting term $\pm L_{\rm A}$, where
\begin{equation}
L_{\rm A}
=
\frac{1}{2}M_{\rm s}
\left(
\ell_{\rm ex1}^{2}
-
\ell_{\rm ex2}^{2}
\right)
\left(
k_x^2-k_z^2
\right).
\label{LA}
\end{equation}
Importantly, the exchange-axis interchange
$\ell^{(\nu)}_{{\rm ex}_x} \leftrightarrow \ell^{(\nu)}_{{\rm ex}_z}$
between neighboring layers constitutes the continuum analogue
of the sublattice rotation operation of crystalline
altermagnets. The synthetic bilayer remains invariant under
a combined $90^\circ$ rotation of the exchange principal axes
and sublattice interchange, under which
$L_A \propto (k_x^2-k_z^2)$ changes sign. Consequently, the synthetic altermagnetic splitting has the characteristic angular dependence $k_x^2-k_z^2$, vanishing along the nodal directions $k_x=\pm k_z$ in the exchange-only limit. Dipolar interactions, retained in the full calculation, hybridize the split branches and generate the residual anticrossings discussed below.
\begin{figure}[t]
\centering
\includegraphics[width=0.99\columnwidth]{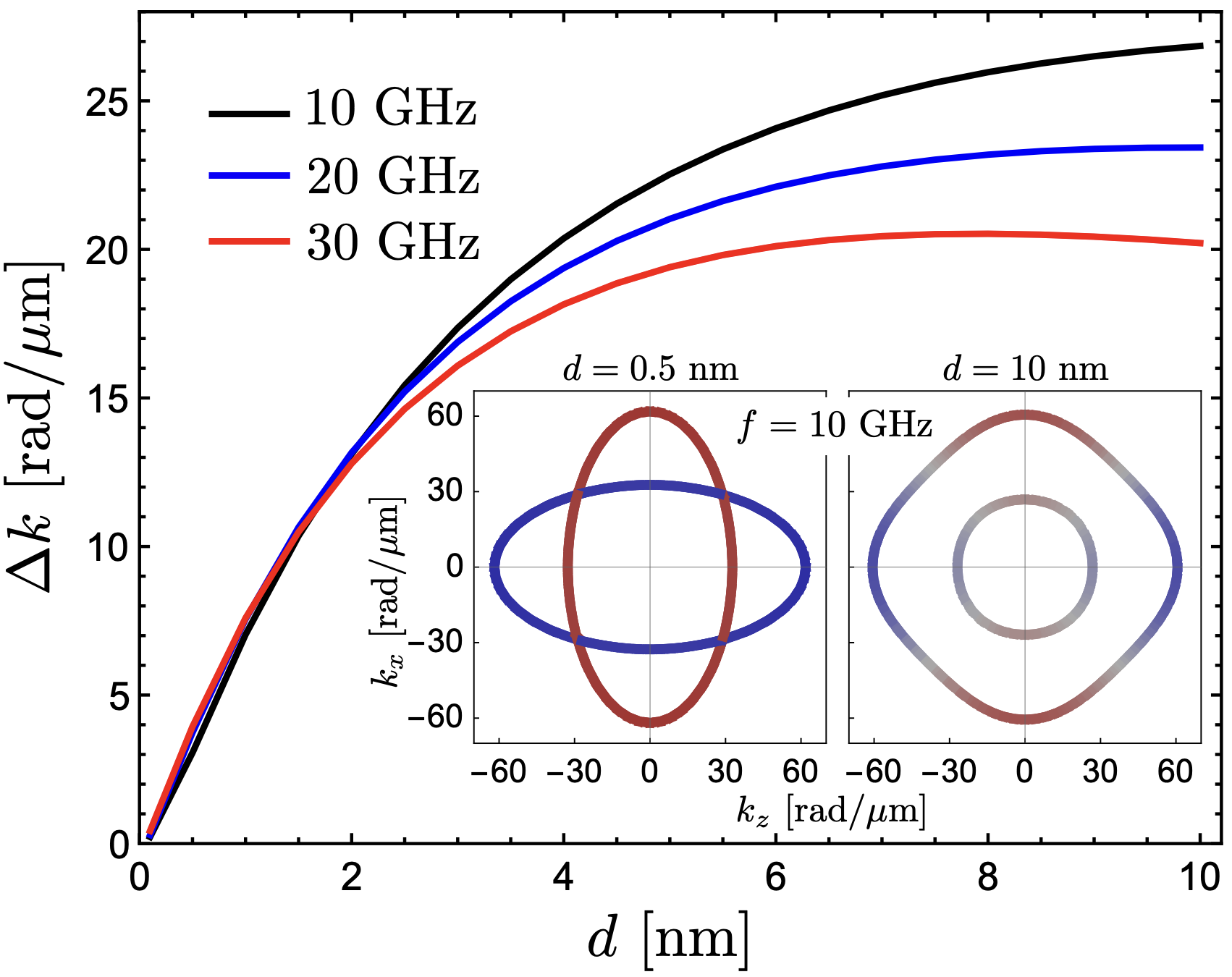}
\caption{
Wave-vector separation $\Delta k$ along the $\Gamma$--$\rm M$ direction as a function of the magnetic layer thickness $d$ for excitation frequencies of 10, 20, and 30 GHz. The increase of $\Delta k$ with thickness reflects the progressive enhancement of magnetostatic effects in the dipole-exchange regime. Insets show the isofrequency contours at $f=10$ GHz for two representative thicknesses, $d=0.5$ nm and $d=10$ nm. As the thickness increases, the contours become increasingly distorted due to the stronger influence of dipolar interactions, leading to an enhanced wave-vector separation along the $\Gamma$--$\rm M$ propagation direction.
}
\label{FIG2}
\end{figure}

For the calculations, we employ representative parameters of PMA-engineered CoFeB/MgO multilayers \cite{Barsukov15,Akyol19}, namely $M_s = 1.0~\mathrm{MA/m}$ and $\gamma/2\pi = 28~\mathrm{GHz/T}$. A perpendicular equilibrium configuration is stabilized by an effective anisotropy field $\mu_0 H_{\perp} = 1.1\,\mu_0 M_s$, representative of interface-induced PMA in CoFeB/MgO heterostructures \cite{Ikeda10,Young14}. To highlight the synthetic altermagnetic splitting, we introduce an engineered exchange anisotropy characterized by $\ell_{{\rm ex}2}/\ell_{{\rm ex}1}=2$, $\ell_{{\rm ex}1}=3~\mathrm{nm}$, and spacer thickness $s=0.5~\mathrm{nm}$. For odd layers,
\begin{equation}
\ell^{(\nu)}_{{\rm ex}_{x}}=\ell_{\rm ex1},
\qquad
\ell^{(\nu)}_{{\rm ex}_{z}}=\ell_{\rm ex2},
\end{equation}
while for even layers $\ell_{\rm ex1}\leftrightarrow\ell_{\rm ex2}$. 
This alternating exchange pattern produces opposite anisotropic exchange
environments on neighboring magnetic sublattices, thereby realizing the conditions required for a synthetic altermagnetic state. 
A plausible experimental route is a W/Co/Pt/Ru/Pt/Co/W stack grown on a (110)-oriented substrate. 
The low-symmetry substrate can break the fourfold in-plane symmetry and induce anisotropic spin-wave stiffness~\cite{Fallarino17,Duine18}, while inversion of the heavy-metal sequence from W/Co/Pt to Pt/Co/W reverses the sign of the interfacial Dzyaloshinskii--Moriya interaction~\cite{Jena21}. 
The Ru spacer supplies strong antiferromagnetic RKKY coupling~\cite{liu2019strong,Duine18}, and the Co/Pt interfaces maintain perpendicular magnetic anisotropy~\cite{ishibashi2021spin}. 
Although the microscopic mechanism differs from the anisotropic-exchange model considered here, such structures realize the same symmetry principle of alternating anisotropic spin-wave environments on neighboring magnetic sublattices and therefore provide a physically plausible route toward synthetic altermagnetism. Additional details on candidate platforms and detection strategies are provided in the Supplemental Material~\cite{SuppMater}.

As a first demonstration, we consider the minimal synthetic altermagnetic realization consisting of two antiferromagnetically coupled ferromagnetic layers ($N=2$) with $J_{\rm I}=-0.1$ mJ/m$^2$ and $s=0.5$ nm. The equilibrium magnetization is stabilized in the perpendicular direction by a uniaxial anisotropy field $\mu_0H_{\perp}=1.1\mu_0M_{\rm s}$.
Since the multilayer is translationally invariant in the film plane, no Brillouin zone exists. Nevertheless, to facilitate the visualization of the angular dependence of the spectrum, we introduce the reference points $\Gamma=(0,0)$, ${\rm X}=(0,k_0)$, ${\rm Y}=(k_0,0)$, and ${\rm M}=(k_0,k_0)$ in wave-vector space, where $k_0=100~{\rm rad/\mu m}$ is chosen to lie within the range accessible to time-resolved x-ray microscopy measurements \cite{Dieterle19,Mayr24}. 
These labels are used only as convenient markers for propagation along the principal and diagonal directions of the $(k_x,k_z)$ plane.
The resulting spin-wave spectrum is shown in Fig.~\ref{FIG1}. The dispersion along the high-symmetry path $\rm Y$--$\Gamma$--$\rm X$--$\rm M$--$\Gamma$ [Fig.~\ref{FIG1}(a)] reveals two nondegenerate magnon branches despite the fully compensated magnetic order. The splitting vanishes at $\Gamma$ and increases with wave vector, reflecting its exchange origin. Moreover, the splitting is strongly direction dependent: it is maximal along the principal axes and substantially reduced along the diagonal direction, consistent with the characteristic altermagnetic symmetry $k_x^2-k_z^2$.

To quantify the sublattice polarization of the $n$th magnon 
mode, we define
\begin{equation}
\chi_n(\mathbf{k})
=
\frac{
\sum_{\nu\in{\rm odd}}
|\mathbf{m}_\nu^{(n)}|^2
-
\sum_{\nu\in{\rm even}}
|\mathbf{m}_\nu^{(n)}|^2
}{
\sum_{\nu=1}^{N}
|\mathbf{m}_\nu^{(n)}|^2
},
\label{eq:chi}
\end{equation}
where $|\mathbf{m}_\nu^{(n)}|^2 = |m_{x_\nu}^{(n)}|^2 + |m_{z_\nu}^{(n)}|^2$ is the spin-wave intensity on layer $\nu$. Here $\chi_n = +1$ and  $\chi_n = -1$ correspond to complete localization on the odd and even magnetic sublattices, respectively, whereas $\chi_n = 0$ indicates equal distribution of the dynamical amplitude between the two sublattices. 
This quantity measures sublattice polarization rather than circular precessional chirality.
The reciprocal-space structure of the modes is visualized through the isofrequency contours shown in Fig.~\ref{FIG1}(b). 
The two branches develop anisotropic contours elongated along orthogonal directions, producing a momentum-dependent separation that increases with frequency. 
The color scale represents $\chi_n(\mathbf{k})$, revealing that the split branches have opposite sublattice polarization. 
The altermagnetic splitting is therefore accompanied by a 
sublattice-polarization splitting, whereby each branch becomes predominantly localized on one magnetic sublattice. 
Along the nominal nodal direction $\Gamma$--M, a finite wave-vector separation $\Delta k$ persists due to dipolar-induced hybridization, which lifts the exchange-only degeneracy expected for $k_x = \pm k_z$.

Further insight is provided by the layer-resolved precessional trajectories shown in Figs.~\ref{FIG1}(c,d). The two modes exhibit distinct phase relations and precessional amplitudes on the magnetic sublattices, reflecting the sublattice-selective character of the synthetic altermagnetic splitting. Consequently, the spectral splitting is accompanied by a redistribution of the dynamical magnetization between the two sublattices, leading to magnon modes with opposite polarization character. These results demonstrate that a simple dipole-exchange bilayer already reproduces the key signatures of altermagnetic magnonics, including momentum-dependent branch splitting, nodal directions, anisotropic isofrequency contours, and sublattice-selective collective dynamics.

The role of dipolar interactions is examined in Fig.~\ref{FIG2} through the wave-vector separation $\Delta k$ measured along the diagonal $\Gamma$--$\rm M$ direction. Unlike the splitting discussed in Fig.~\ref{FIG1}, which originates from the exchange symmetry term $k_x^2-k_z^2$, $\Delta k$ probes the residual branch separation where the exchange contribution vanishes.
Figure~\ref{FIG2}(a) shows that $\Delta k$ increases monotonically with magnetic-layer thickness, demonstrating the growing influence of magnetostatic coupling. The effect is most pronounced at low frequencies, where dipolar interactions dominate the collective dynamics, and progressively weakens as the spectrum becomes more exchange dominated. The corresponding isofrequency contours [Figs.~\ref{FIG2}(b,c)] reveal a transition from weakly perturbed exchange-driven contours in ultrathin layers to strongly distorted contours exhibiting a clear separation along the diagonal direction. These results identify dipolar coupling as the origin of the finite nodal splitting observed in the synthetic altermagnetic spectrum.

The physical origin of the synthetic altermagnetic splitting can be understood from the analytical bilayer dispersion derived in the Supplemental Material \cite{SuppMater}. In the exchange-only limit ($d\rightarrow0$), the dipolar contributions vanish and the eigenfrequencies reduce to
\begin{equation}
f
=
\frac{\mu_0\gamma}{2\pi}
\left[
\sqrt{L_{\mathbf k}
\left(
L_{\mathbf k}- 2 M_{\rm s}c_{\rm J}
\right)}
\pm
L_{\rm A}
\right].
\end{equation}
The splitting is therefore governed entirely by $L_{\rm A}$ [see Eq.~\eqref{LA}] and vanishes along the nodal directions $k_x=\pm k_z$, reproducing the characteristic momentum-space symmetry of altermagnetic systems \cite{Libor22}.
When dipolar interactions are included, the nodal degeneracy is lifted. For propagation along $\Gamma$--$\rm M$ ($L_{\rm A}=0$) and in the ultrathin limit $d|\mathbf{k}|\ll1$, and with $c_{\rm J}=J_{\rm I}/(\mu_0 dM_{\rm s}^2)$ the dispersion simplifies to
\begin{equation}
f_{\pm}\simeq
\frac{\mu_0\gamma}{2\pi}
\sqrt{
L_{\mathbf k}
\left[
L_{\mathbf k}
+
M_{\rm s}
\left(
\frac{d|\mathbf{k}|}{2}
\pm
\frac{d|\mathbf{k}|}{2}
-
2c_{\rm J}
\right)
\right]
},
\end{equation}
showing explicitly that dipolar interactions generate a finite branch separation even along the nominal nodal direction. The synthetic altermagnetic response therefore results from two complementary mechanisms: an exchange-driven splitting with $d$-wave symmetry and a dipolar-induced hybridization that produces finite avoided crossings at the nodal points.

\begin{figure*}[t]
\includegraphics[width=0.99\textwidth]{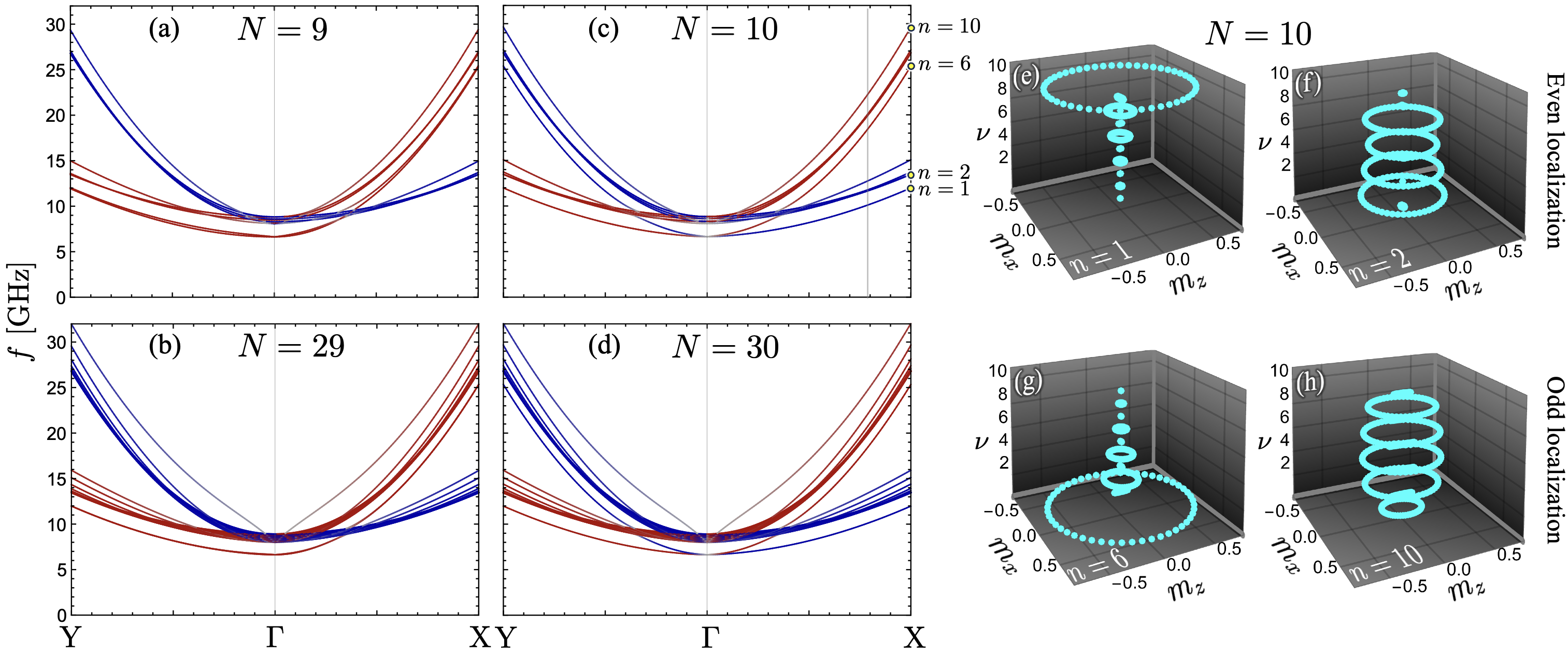}
\caption{
(a,b) Spin-wave dispersions for synthetic altermagnetic multilayers with odd numbers of magnetic layers, $N=9$ and $N=29$, respectively. (c,d) Corresponding spectra for even multilayers with $N=10$ and $N=30$. 
(e--h) Layer-resolved precessional trajectories of selected modes marked in panel (c) (modes $n=1$, 2, 6 and 10 evaluated at the point $\rm X$), represented through the normalized dynamical magnetization components $(m_x,m_z)$ as a function of the layer index $\nu$. The even and odd magnetization oscillations illustrate the sublattice-selective dynamical response that emerges in the synthetic altermagnetic phase.
}
  \label{FIG3}
\end{figure*}

Figure~\ref{FIG3} shows the evolution of the synthetic altermagnetic response as the bilayer is extended into a multilayer structure. Increasing the number of magnetic layers transforms the spectrum into a dense multiband system composed of surface-localized and bulk collective excitations. Beyond the bilayer limit, a parity-dependent reconstruction of the altermagnetic response emerges. For even multilayers, both surface and bulk modes preserve the characteristic altermagnetic branch splitting [Figs.~\ref{FIG3}(c,d)], whereas in odd multilayers the surface modes progressively lose this character while the bulk modes remain split [Figs.~\ref{FIG3}(a,b)]. This behavior originates from the uncompensated termination of odd stacks, which breaks the equivalence between the two magnetic sublattices at the surfaces. Consequently, synthetic altermagnetism develops a distinct surface--bulk dichotomy that is absent in the bilayer limit.

The layer-resolved trajectories in Figs.~\ref{FIG3}(e)--\ref{FIG3}(h), evaluated at the $\rm X$ point, illustrate the dynamical origin of this behavior for $N=10$. While the surface modes are localized near opposite boundaries of the stack, the bulk modes extend throughout the multilayer and develop a pronounced sublattice selectivity, with the precessional amplitude concentrated predominantly on either the odd or even magnetic sublattice, consistent with the polarization parameter $\chi_n(\mathbf{k})$. These profiles demonstrate that the momentum-dependent altermagnetic splitting is accompanied by a polarization splitting of the collective dynamics. As a result, the multilayer supports bulk excitations that simultaneously exhibit altermagnetic branch splitting and sublattice-selective dynamics, establishing synthetic altermagnetism as a spatially structured collective response rather than a uniform property of the entire spectrum.

In conclusion, we demonstrate a synthetic realization of an altermagnetic system composed by antiferromagnetically coupled ferromagnetic multilayers with alternating in-plane exchange anisotropies. 
The alternating exchange environment generates a continuum altermagnetic term proportional to $k_x^2-k_z^2$, reproducing the momentum-dependent splitting, nodal directions, and anisotropic isofrequency contours characteristic of A-type altermagnets. 
Within a full dipole-exchange treatment, we show that long-range magnetostatic interactions qualitatively reconstruct this exchange-driven spectrum by hybridizing opposite-chirality modes, lifting the nominal nodal degeneracy, and producing finite, thickness-dependent wave-vector splitting and avoided crossings along directions that are nodal in the exchange-only limit. 
Finite multilayers reveal that synthetic altermagnetism is not a uniform property of the spectrum but undergoes a parity-dependent reconstruction that separates surface and bulk altermagnetic excitations. 
These results establish altermagnetic magnon phenomenology as an engineerable collective response in dipole-exchange multilayers that goes beyond microscopic crystal symmetries.
The parity and surface-termination effects identified here emerge as general design principles for synthetic altermagnetism in vertically stacked heterostructures. 
Their direct analogue in oxide superlattices, where even-numbered magnetic blocks and sufficient spacer decoupling are required to sustain the altermagnetic state~\cite{Xiang26}, suggests that these design principles may be largely independent of the microscopic mechanism, whether exchange anisotropy with dipolar coupling or interfacial orbital reconstruction with quantum confinement. 
The proposed platform is experimentally accessible in perpendicularly magnetized metallic multilayers with RKKY antiferromagnetic coupling, where the predicted altermagnetic spin-wave branches are resolvable via Mie-enhanced microfocused Brillouin light scattering~\cite{krcma2025mie} (see Supplemental Material~\cite{SuppMater} for candidate platforms and detection strategies). 
Importantly, this realization demonstrates that altermagnetic magnonic phenomena can emerge from simple magnetic building blocks rather than requiring intrinsically altermagnetic compounds.
More broadly, the continuum dipole-exchange framework developed here applies to other A-type altermagnetic configurations and opens new routes for engineering chiral magnon dynamics in synthetic compensated magnets.

\section*{ACKNOWLEDGMENTS}

The authors acknowledge financial support from the Fondecyt grant 1250803 and CEDENNA under grant CIA250002 from ANID. A.L. acknowledges support from ANID through FONDECYT Iniciación Grant No. 11251906.

\sloppy
\bibliography{AMBIB}

\clearpage
\onecolumngrid

\begin{center}
{\Large \textbf{Supplemental Material}}\\[0.3cm]
{\large \textbf{Synthetic Altermagnetism Beyond the Crystal Limit}}
\end{center}

\setcounter{section}{0}
\renewcommand{\thesection}{S\arabic{section}}

\setcounter{equation}{0}
\renewcommand{\theequation}{S\arabic{equation}}

\setcounter{figure}{0}
\renewcommand{\thefigure}{S\arabic{figure}}

\setcounter{table}{0}
\renewcommand{\thetable}{S\arabic{table}}

\section{Theoretical model and dynamical matrix}
\label{App:Matrix}

\begin{figure}[t]
\centering
\includegraphics[width=0.5\columnwidth]{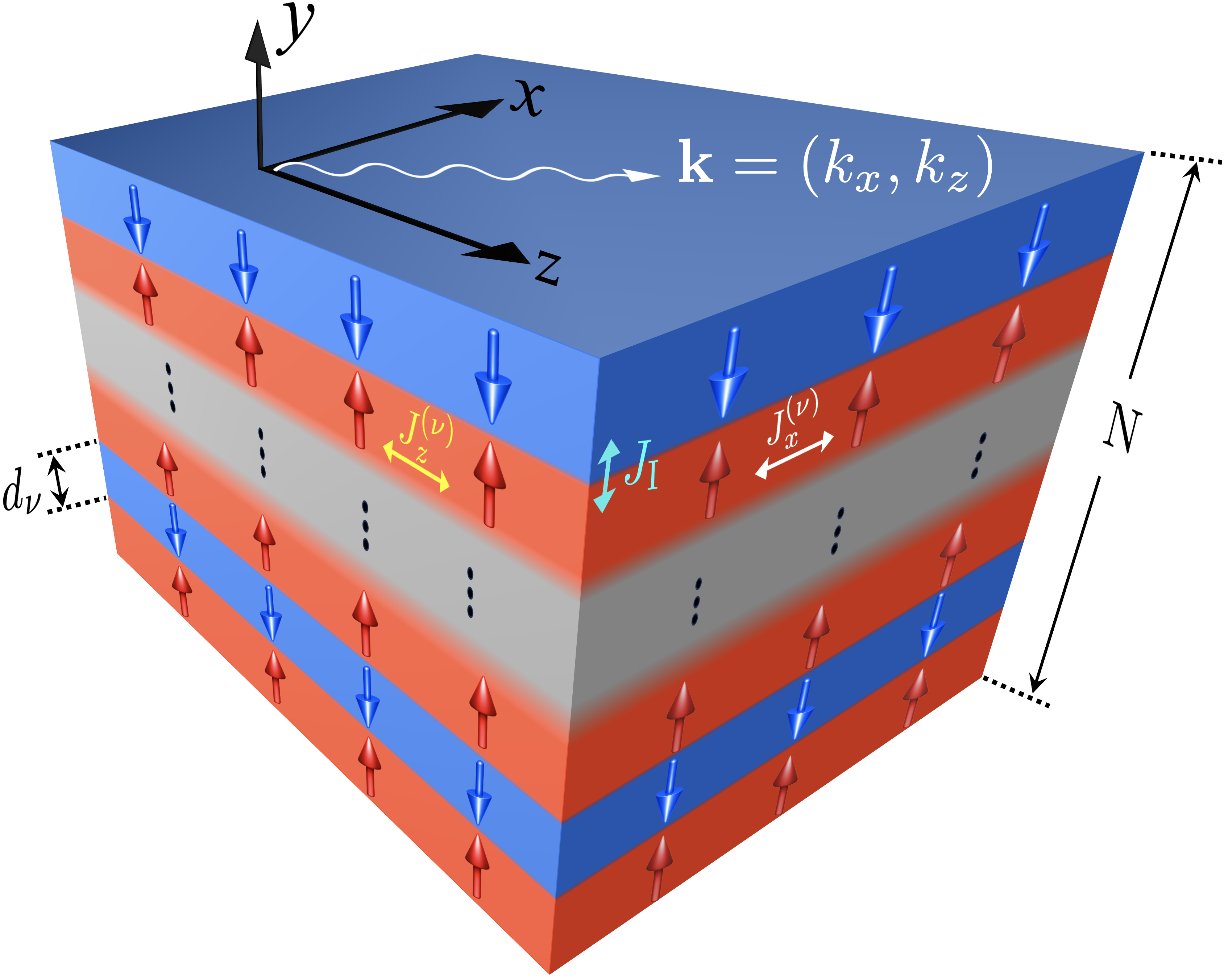}
\caption{
Schematic illustration of the synthetic altermagnetic multilayer system. Ferromagnetic layers with alternating out-of-plane magnetization (up, blue; down, red) are antiferromagnetically coupled through the interlayer exchange interaction ($J_{\mathrm{I}}<0$). The in-plane exchange interaction is anisotropic, with different exchange strengths along the $x$ [$J_x^{(\nu)}$] and $z$ [$J_z^{(\nu)}$] directions. Spin waves propagate along an arbitrary in-plane wave vector $\mathbf{k}=(k_x,k_z)$. The total number of layers is denoted by $N$, while $d_{\nu}$ represents the thickness of the $\nu$th magnetic layer.
}
\label{FIGS1}
\end{figure}

We consider a synthetic altermagnetic multilayer composed of $N$ ferromagnetic films antiferromagnetically coupled along the stacking direction, as schematically illustrated in Fig.~\ref{FIGS1}. The multilayer is laterally extended in the $x$--$z$ plane, while the layer normal is oriented along the $y$ direction. Each magnetic layer is assumed to be uniformly magnetized in equilibrium along the perpendicular direction, with
\begin{equation}
\mathbf{M}_{\nu}^{\mathrm{eq}}
=
(-1)^{\nu+1}M_{\mathrm{s}}\hat{\mathbf{y}},
\end{equation}
where $\nu=1,\ldots,N$ labels the magnetic layers. Thus, the bottom layer ($\nu=1$) is oriented along $+y$, while neighboring layers alternate their equilibrium magnetization direction due to the antiferromagnetic interlayer exchange interaction.
The magnetization dynamics in the $\nu$th layer is described by the Landau--Lifshitz equation
\begin{equation}
\frac{d}{dt}\mathbf{M}_{\nu}(\mathbf{r},t)
=
-\mu_0 \gamma\,
\mathbf{M}_{\nu}(\mathbf{r},t)
\times
\mathbf{H}_{\nu}^{\mathrm{eff}}(\mathbf{r},t),
\label{EQ1SM}
\end{equation}
where $\gamma$ is the gyromagnetic ratio and $\mathbf{H}_{\nu}^{\mathrm{eff}}$ is the effective field acting on the $\nu$th layer. Spin-wave excitations propagating with an arbitrary in-plane wave vector $\mathbf{k}=(k_x,k_z)$ are introduced through
\begin{equation}
\mathbf{M}_{\nu}(\mathbf{r},t)
=
\mathbf{M}_{\nu}^{\mathrm{eq}}
+
\mathbf{m}_{\nu}
e^{i(k_xx+k_zz-\omega t)},
\end{equation}
where $\mathbf{m}_{\nu}$ is the dynamic magnetization amplitude. The effective field is decomposed into static and dynamic contributions as
\begin{equation}
\mathbf{H}_{\nu}^{\mathrm{eff}}(\mathbf{r},t)
=
\mathbf{H}_{\nu}^{\mathrm{eq}}
+
\mathbf{h}_{\nu}
e^{i(k_xx+k_zz-\omega t)}.
\end{equation}
Keeping only terms linear in $\mathbf{m}_{\nu}$, Eq.~\eqref{EQ1SM} becomes
\begin{equation}
\omega \mathbf{m}_{\nu}
=
i\gamma\mu_0
\left(
\mathbf{H}_{\nu}^{\mathrm{eq}}
\times
\mathbf{m}_{\nu}
+
\mathbf{h}_{\nu}
\times
\mathbf{M}_{\nu}^{\mathrm{eq}}
\right).
\label{EqlinSM}
\end{equation}

The effective field includes intralayer exchange, perpendicular anisotropy, interlayer exchange, and dipolar contributions. The synthetic altermagnetic character is introduced through an anisotropic intralayer exchange interaction whose principal axes are interchanged between neighboring magnetic layers. Within the continuum approximation, this contribution is written as
\begin{equation}
\mathbf{H}_{\nu}^{\mathrm{ex}}
=
\left[\ell^{(\nu)}_{{\rm ex}_x}\right]^2
\frac{\partial^2\mathbf{M}_{\nu}}{\partial x^2}
+
\left[\ell^{(\nu)}_{{\rm ex}_z}\right]^2
\frac{\partial^2\mathbf{M}_{\nu}}{\partial z^2},
\label{HexSM}
\end{equation}
where $\ell^{(\nu)}_{{\rm ex}_x}$ and $\ell^{(\nu)}_{{\rm ex}_z}$ are the exchange lengths along the in-plane $x$ and $z$ directions, respectively. The layer dependence of these exchange lengths encodes the alternating anisotropic exchange environment responsible for the synthetic altermagnetic response.

Adjacent magnetic layers are coupled through an interfacial 
exchange energy per unit area
\begin{equation}
\mathcal{E}_{\nu\eta}^{\mathrm{I}}
=
-
\frac{J_{\mathrm{I}}}{M_{\mathrm{s}}^2}
\mathbf{M}_{\nu}\cdot\mathbf{M}_{\eta},
\end{equation}
where $J_{\mathrm{I}}<0$ stabilizes the antiferromagnetic 
alignment. Averaging over the layer thickness $d_\nu$ yields 
the volumetric energy density, from which the corresponding 
effective field acting on layer $\nu$ is obtained as
\begin{equation}
\mathbf{H}_{\nu}^{\mathrm{I}}
=
\sum_{\eta}
\frac{J_{\mathrm{I}}}
{\mu_0 d_{\nu}M_{\mathrm{s}}^2}
\mathbf{M}_{\eta},
\label{HI_SM}
\end{equation}
where $d_{\nu}$ is the thickness of the $\nu$th magnetic layer 
and the sum extends over the nearest-neighbor magnetic layers 
coupled to layer $\nu$.

The dipolar interaction is treated within the full magnetostatic formalism following the approach of Ref.~\onlinecite{Contreras25}, generalized here to the perpendicular equilibrium configuration.

The dynamic components of the effective field can be written in the compact form
\begin{equation}
h_{\alpha_\nu}^{\rm eff}(\mathbf{r})
=
\sum_{\eta=1}^{N}
\sum_{\alpha'=x,z}
\Gamma_{\alpha_\nu\alpha'_\eta}
m_{\alpha'_{\eta}}(\mathbf{r}),
\label{effectivefieldSM}
\end{equation}
where $\alpha,\alpha'=x,z$. Inserting Eq.~\eqref{effectivefieldSM} into Eq.~\eqref{EqlinSM}, the linearized equations can be cast as the eigenvalue problem
\begin{equation}
i\frac{\omega}{\mu_0\gamma}\mathbf{m}
=
\widetilde{\mathcal{D}}\mathbf{m},
\label{EQavSM}
\end{equation}
where $\mathbf{m}=(m_{x_1},m_{z_1},\ldots,m_{x_N},m_{z_N})^{T}$ contains the dynamic magnetization amplitudes of all magnetic layers. The matrix $\widetilde{\mathcal{D}}$ has dimension $2N\times2N$. Its elements are
\begin{eqnarray}
\mathcal{D}_{zz}^{\nu\eta}
&=&
M_{\rm s}\Gamma_{xz}^{\nu\eta},
\\
\mathcal{D}_{zx}^{\nu\eta}
&=&
M_{\rm s}\Gamma_{xx}^{\nu\eta}
-
H^{\rm eq}_{y_\nu}\delta_{\nu}^{\eta},
\\
\mathcal{D}_{xz}^{\nu\eta}
&=&
H^{\rm eq}_{y_\nu}\delta_{\nu}^{\eta}
-
M_{\rm s}\Gamma_{zz}^{\nu\eta},
\\
\mathcal{D}_{xx}^{\nu\eta}
&=&
-
M_{\rm s}\Gamma_{zx}^{\nu\eta}.
\end{eqnarray}
Here, $\delta_{\nu}^{\eta}$ is the Kronecker delta and $H^{\rm eq}_{y_\nu}$ is the static perpendicular component of the effective field in the $\nu$th layer.
Defining $|\mathbf{k}|=\sqrt{k_x^2+k_z^2}$ and
\begin{equation}
c_{\rm J}=
\frac{J_{\rm I}}{\mu_0 d_{\nu}M_{\rm s}^2},
\end{equation}
the coefficients $\Gamma_{\alpha\alpha'}^{\nu\eta}$ read
\begin{equation}
\Gamma_{xz}^{\nu\eta}=\Gamma_{zx}^{\nu\eta}
=
\frac{k_x k_z(-1)^{\eta+1}}{|\mathbf{k}|^2}
\left[
\mathcal{P}^{\nu\eta}_{\mathbf{k}}
(1-\delta^{\eta}_{\nu})
+
\zeta^{\nu}_{\mathbf{k}}\delta^{\eta}_{\nu}
\right],
\end{equation}
\begin{equation}
\Gamma_{xx}^{\nu\eta}
=
(-1)^{\eta+\nu}\left[c_{\rm J}
\left(
\delta^{\eta}_{\nu-1}
+
\delta^{\eta}_{\nu+1}
\right)
-
\frac{k_x^2 \mathcal{P}^{\nu\eta}_{\mathbf{k}}}{|\mathbf{k}|^2}
(1-\delta^{\eta}_{\nu})\right]-
\left[
k_x^2\left(\ell_{{\rm ex}_x}^{(\eta)}\right)^2
+
k_z^2\left(\ell_{{\rm ex}_z}^{(\eta)}\right)^2
+
\frac{k_x^2\zeta^{\nu}_{\mathbf{k}}}{|\mathbf{k}|^2}
\right]\delta^{\eta}_{\nu},
\end{equation}
and
\begin{equation}
\Gamma_{zz}^{\nu\eta}
=
c_{\rm J}
\left(
\delta^{\eta}_{\nu-1}
+
\delta^{\eta}_{\nu+1}
\right)
-
\frac{k_z^2\mathcal{P}^{\nu\eta}_{\mathbf{k}}}{|\mathbf{k}|^2}
(1-\delta^{\eta}_{\nu})-
\left[
k_x^2\left(\ell_{{\rm ex}_x}^{(\eta)}\right)^2
+
k_z^2\left(\ell_{{\rm ex}_z}^{(\eta)}\right)^2
+
\frac{k_z^2\zeta^{\nu}_{\mathbf{k}}}{|\mathbf{k}|^2}
\right]\delta^{\eta}_{\nu},
\end{equation}

The terms proportional to $\delta^{\eta}_{\nu}$ describe intralayer contributions, whereas those proportional to $1-\delta^{\eta}_{\nu}$ and $\delta^{\eta}_{\nu\pm1}$ describe interlayer dipolar and exchange contributions, respectively.
The dimensionless functions $\zeta^{\nu}_{\mathbf{k}}$ and $\mathcal{P}^{\nu\eta}_{\mathbf{k}}$ originate from the intralayer and interlayer dipolar interactions and are given by
\begin{equation}
\zeta^{\nu}_{\mathbf{k}}
=
1-
\frac{
2\sinh\left(|\mathbf{k}|d_{\nu}/2\right)
}{
d_{\nu}|\mathbf{k}|
}
e^{-|\mathbf{k}|d_{\nu}/2},
\end{equation}
and
\begin{equation}
\mathcal{P}^{\nu\eta}_{\mathbf{k}}
=
\frac{
2
\sinh\left(|\mathbf{k}|d_{\eta}/2\right)
\sinh\left(|\mathbf{k}|d_{\nu}/2\right)
}{
d_{\eta}|\mathbf{k}|
}
e^{-|\mathbf{k}||\xi_{\eta}-\xi_{\nu}|},
\end{equation}
with
\begin{equation}
\xi_{\eta}
=
\sum_{j=1}^{\eta-1}(d_j+s_j)
+
\frac{d_{\eta}}{2},
\end{equation}
where $s_j$ is the nonmagnetic spacer thickness between neighboring magnetic layers. Although the formalism allows for layer-dependent thicknesses, all magnetic layers are taken to have the same thickness in the calculations reported in the main text.

The static perpendicular field entering the dynamical matrix is
\begin{equation}
H^{\rm eq}_{y_\nu}
=
M_{\rm s}c_{\rm J}
\sum_{\eta=1}^{N}
\left(
\delta^{\eta}_{\nu-1}
+
\delta^{\eta}_{\nu+1}
\right)
(-1)^{\eta+\nu}
+
H_{\perp}
-
M_{\rm s},
\end{equation}
where $H_{\perp}$ is the perpendicular anisotropy field. This term stabilizes the out-of-plane equilibrium state and may arise from interfacial anisotropy at the ferromagnet/nonmagnetic interfaces.

The numerical solution of Eq.~\eqref{EQavSM} yields the spin-wave frequencies $f(\mathbf{k})=\omega(\mathbf{k})/2\pi$ and the corresponding eigenvectors, from which the layer-resolved mode profiles, surface localization, polarization character, and isofrequency contours discussed in the main text are obtained.

\section{Analytical bilayer dispersion}
\label{App:MatrixB}

For the bilayer system ($N=2$), the two magnetic layers have identical thicknesses, $d_1=d_2=d$, so that
\begin{equation}
\zeta_{\mathbf{k}}^{1}
=
\zeta_{\mathbf{k}}^{2}
\equiv
\zeta_{\mathbf{k}}.
\end{equation}
Likewise,
\begin{equation}
\mathcal{P}_{\mathbf{k}}^{12}
=
\mathcal{P}_{\mathbf{k}}^{21}
\equiv
\mathcal{P}_{\mathbf{k}},
\end{equation}
with $|\xi_1-\xi_2|=d+s$, where $s$ is the spacer thickness. After algebraic manipulation of the resulting $4\times4$ dynamical matrix, the bilayer dispersion relation can be written as
\begin{equation}
f
=
\frac{\mu_0\gamma}{2\pi}
\sqrt{
\frac{1}{2}
\left(
-\mathcal{B}
\pm
\sqrt{\mathcal{B}^2-4\mathcal{A}}
\right)
},
\label{SWfullSM}
\end{equation}
where
\begin{equation}
\mathcal{A}
=
\left[
L_{\mathbf{k}}
\left(
L_{\mathbf{k}}-2M_{\rm s}c_{\rm J}
\right)
-
L_{\rm A}^2
\right]
\left[
\Lambda_{-}\Lambda_{+}
-
L_{\rm A}^2
-
2M_{\rm s}c_{\rm J}\Lambda_{-}
\right].
\end{equation}
\begin{equation}
\mathcal{B}
=
-2
\left[
L_{\mathbf{k}}
\left(
L_{\mathbf{k}}
+
M_{\rm s}\zeta_{\mathbf{k}}
-
2M_{\rm s}c_{\rm J}
\right)
+
L_{\rm A}^2
-
M_{\rm s}^{2}c_{\rm J}
(\zeta_{\mathbf{k}}-\mathcal{P}_{\mathbf{k}})
\right].
\end{equation}

Here,
\begin{equation}
\Lambda_{\pm}=L_{\mathbf{k}}+M_{\rm s}
\left(\zeta_{\mathbf{k}}\pm\mathcal{P}_{\mathbf{k}}\right),
\end{equation}
\begin{equation}
L_{\mathbf{k}}
=
H_{\perp}
-
M_{\rm s}
+
\frac{1}{2}M_{\rm s}
\left(
\ell_{\rm ex1}^{2}
+
\ell_{\rm ex2}^{2}
\right)
\left(
k_x^2+k_z^2
\right),
\end{equation}
and
\begin{equation}
L_{\rm A}
=
\frac{1}{2}M_{\rm s}
\left(
\ell_{\rm ex1}^{2}
-
\ell_{\rm ex2}^{2}
\right)
\left(
k_x^2-k_z^2
\right).
\end{equation}
The term $L_{\mathbf{k}}$ contains the isotropic part of the exchange stiffness, whereas $L_{\rm A}$ contains the anisotropic contribution that changes sign under the interchange $k_x\leftrightarrow k_z$. Therefore, $L_{\rm A}$ is the exchange-origin term responsible for the characteristic altermagnetic-like angular dependence of the magnon splitting.

Equation~\eqref{SWfullSM} includes both anisotropic exchange and dipole-mediated contributions. The two signs in Eq.~\eqref{SWfullSM} give the two bilayer magnon branches. In the absence of dipolar interactions, the splitting is controlled by $L_{\rm A}$ and therefore vanishes along the nodal directions $k_x=\pm k_z$. When dipolar interactions are retained, the functions $\zeta_{\mathbf{k}}$ and $\mathcal{P}_{\mathbf{k}}$ introduce additional magnetostatic hybridization between the branches, lifting the purely exchange nodal degeneracy and producing the residual anticrossing discussed in the main text.

The synthetic altermagnetic response is robust with respect to the magnitude of the exchange anisotropy. The branch splitting vanishes continuously as $\ell_{ex2}/\ell_{ex1} \to 1$ and appears for any finite deviation from an isotropic exchange environment. The value $\ell_{ex2}/\ell_{ex1} = 2$ used in the main text is not a threshold condition but a representative choice that makes the splitting clearly visible in the figures. Smaller exchange-anisotropy ratios preserve the same $d$-wave angular dependence and nodal structure, with a proportionally reduced splitting amplitude.

\section{Candidate Experimental Platforms}

Several experimental platforms provide the essential ingredients for realizing the proposed synthetic altermagnetic multilayer: perpendicular magnetic anisotropy, antiferromagnetic interlayer exchange coupling, and a sublattice-dependent in-plane exchange environment. Perpendicularly magnetized $[\mathrm{Co/Pt}]_n/\mathrm{Ru}/[\mathrm{Co/Pt}]_n$ structures satisfy the first two requirements, with interlayer exchange constants reaching $J_{\rm I}\approx -1.77$~mJ/m$^2$ at the first RKKY peak for $t_{\rm Ru}\approx0.44$~nm~\cite{liu2019strong}. Even stronger antiferromagnetic coupling has been demonstrated in $[\mathrm{Co/Pt}]_n/\mathrm{Ir}/[\mathrm{Co/Pt}]_n$ multilayers, where $J_{\rm I}$ can exceed $-2.7$~mJ/m$^2$ \cite{Gabor17}, providing an alternative platform for robust synthetic antiferromagnetic coupling. Similarly, $[\mathrm{Co/Ni}]_n/\mathrm{Ru}/[\mathrm{Co/Ni}]_n$ multilayers with perpendicular anisotropy support antiferromagnetic spin-wave resonance with opposite-chirality modes~\cite{ishibashi2021spin}, in direct analogy with the altermagnetic branches identified in the present work. For systems based on CoFeB, $\mathrm{CoFeB}/\mathrm{Ru}/\mathrm{CoFeB}$ synthetic antiferromagnets offer well-controlled exchange stiffness ($A=16\pm2$~pJ/m) and tunable interlayer coupling~\cite{mouhoub2023exchange}; combining such structures with interfacial PMA layers such as $\mathrm{MgO}$ enables stabilization of the out-of-plane magnetic configuration required here. The short-wavelength altermagnetic spin-wave branches predicted in this work are experimentally accessible using Mie-enhanced microfocused Brillouin light scattering~\cite{grassi2020shortwavelength,krcma2025mie}, as well as scanning transmission x-ray microscopy with wave-vector sensitivity down to tens of nanometers~\cite{Mayr24}. In both approaches, wave-vector-resolved measurements can be performed without significantly perturbing the magnetic properties of the multilayer.

The realization of the synthetic altermagnetic state involves two distinct requirements. First, each magnetic block must exhibit an in-plane anisotropic spin-wave stiffness arising from reduced crystal symmetry. Second, the principal axes of this anisotropy must be interchanged between neighboring magnetic blocks.

The first requirement is experimentally well established. The exchange stiffness tensor of a ferromagnetic layer becomes anisotropic whenever the local crystal symmetry is reduced below fourfold. Epitaxial growth along low-symmetry crystallographic orientations, including Fe(110), Co(110), and Ni(110), naturally produces anisotropic magnetic interactions and direction-dependent spin-wave propagation with principal axes fixed by the crystal lattice~\cite{Duine18,Fallarino17,Tovar93}. Similar effects can be induced through biaxial strain engineering using orthorhombic substrates such as $\mathrm{NdGaO}_3$ or $\mathrm{DyScO}_3$, or through buffer layers that break the in-plane rotational symmetry~\cite{Kirby10,Fallarino17}. Compositional tuning in alloy systems such as $\mathrm{Co}_x\mathrm{Fe}_{1-x}$ and $\mathrm{Ni}_x\mathrm{Fe}_{1-x}$ further modifies the magnitude and anisotropy of the effective spin-wave stiffness~\cite{mouhoub2023exchange}. In addition, ion implantation has been demonstrated as a versatile route for locally tailoring magnetic interactions while preserving the overall multilayer structure and magnetic order~\cite{Gallardo14,Obry13}.

The second requirement, namely the interchange of the principal anisotropy axes between neighboring magnetic blocks, represents the main materials-design challenge. In practice, this may be achieved through layer-selective growth conditions, alternating strain environments, engineered crystallographic texture, or other forms of structural modulation that rotate the local anisotropy axes from one magnetic block to the next. Importantly, this design parameter is independent of the antiferromagnetic interlayer exchange coupling, which is controlled separately through the RKKY spacer thickness~\cite{liu2019strong,Duine18}. The two ingredients can therefore be optimized independently, providing considerable flexibility for the realization of synthetic altermagnetic multilayers.

A complementary realization is provided by $(\mathrm{RuO}_2)_m/(\mathrm{TiO}_2)_n$ oxide superlattices, where interfacial Ti--O--Ru bonding drives orbital reconstruction and establishes N\'eel-type antiferromagnetic order in two-dimensional $\mathrm{RuO}_2$ layers isolated by nonmagnetic $\mathrm{TiO}_2$ spacers~\cite{Xiang26}. The resulting band structure satisfies $E(\mathbf{k},\uparrow)=E(C_{4z}\mathbf{k},\downarrow)$ with spin-degenerate nodal lines along $\Gamma$--X--M, consistent with A-type altermagnetic symmetry. Strikingly, an even number of Ru layers is required to preserve the N\'eel-type sublattice connectivity~\cite{Xiang26}, in direct analogy with the parity-dependent mode reconstruction identified in the present work: even multilayers preserve the altermagnetic branch splitting across both surface and bulk modes, whereas odd multilayers exhibit partial suppression at the surface due to uncompensated sublattice termination. The role of the spacer is equally analogous. A $\mathrm{TiO}_2$ spacer of only two layers is insufficient to fully decouple neighboring $\mathrm{RuO}_2$ blocks, and the altermagnetic state emerges only beyond a critical spacer thickness, mirroring the dependence of dipolar hybridization on the nonmagnetic spacer thickness in the metallic multilayers considered here~\cite{Xiang26}. These parallels suggest that even/odd layer parity and adequate spacer-mediated decoupling may constitute general design principles for synthetic altermagnetism in vertically stacked heterostructures, independent of whether the underlying mechanism originates from exchange anisotropy and dipolar coupling or from interfacial orbital reconstruction and quantum confinement.

A particularly appealing route toward realizing the required layer-dependent anisotropic spin-wave environment exploits interfacial Dzyaloshinskii--Moriya interaction (DMI). In ultrathin heavy-metal/ferromagnet multilayers, the DMI strength and sign are determined by the stacking sequence of the adjacent interfaces. For example, Co layers sandwiched between W and Pt exhibit a DMI whose sign reverses when the order of the heavy-metal layers is inverted, i.e., W/Co/Pt and Pt/Co/W generate opposite chiral interactions. When combined with growth on a low-symmetry substrate, such as a (110)-oriented crystal surface, the interfacial DMI can become anisotropic within the film plane, producing direction-dependent spin-wave propagation. Consequently, a multilayer stack of the form (110) substrate $\rightarrow
\mathrm{W/Co/Pt}\rightarrow
\mathrm{Ru}\rightarrow
\mathrm{Pt/Co/W} \rightarrow$ caplayer 
naturally combines strong antiferromagnetic RKKY coupling with alternating anisotropic chiral interactions in neighboring magnetic layers. Although the microscopic origin differs from the anisotropic exchange model considered in the present work, both systems realize the same essential symmetry principle: neighboring magnetic sublattices experience interchanged anisotropic spin-wave environments. Such structures therefore represent a realistic experimental platform for exploring synthetic altermagnetic magnonics beyond the specific exchange-anisotropy mechanism analyzed here~\cite{Jena21}.

\end{document}